\RequirePackage{fix-cm}
\documentclass[smallextended]{svjour3}       
\smartqed  
\usepackage{graphicx}
\usepackage{program}
\usepackage{cite}
\usepackage{hyperref}
\begin{document}

\title{hashkat: Large-scale simulations of online social networks
}


\author{Kevin Ryczko         \and
        Adam Domurad \and 
        Nicholas Buhagiar \and
        Isaac Tamblyn 
}


\institute{K. Ryczko\at
              University of Ontario Institute of Technology\\
              \email{kevin.ryczko@uoit.net}           
           \and
           A. Domurad \at
              University of Waterloo
           \and
           N. Buhagiar \at
           	  Ryerson University
           \and
           I. Tamblyn \at
             National Research Council of Canada\\
             University of Ontario Institute of Technology\\
             \email{isaac.tamblyn@nrc-cnrc.gc.ca}
}

\date{Received: date / Accepted: date}

\maketitle

\begin{abstract}
Hashkat (\href{http://hashkat.org}{http://hashkat.org}) is a free, open source, agent based simulation software package designed to simulate large-scale online social networks (e.g. Twitter, Facebook, LinkedIn, etc). It allows for dynamic agent generation, edge creation, and information propagation. The purpose of hashkat is to study the growth of online social networks and how information flows within them. Like real life online social networks, hashkat incorporates user relationships, information diffusion, and trending topics. Hashkat was implemented in C++, and was designed with extensibility in mind. The software includes Shell and Python scripts for easy installation and usability. In this report, we describe all of the algorithms and features integrated into hashkat before moving on to example use cases. In general, hashkat can be used to understand the underlying topology of social networks, validate sampling methods of such networks, develop business strategy for advertising on online social networks, and test new features of an online social network before going into production.
\keywords{agent based modelling \and kinetic Monte Carlo \and online social network \and network evolution \and information propagation \and simulation}

\end{abstract}

\section{Introduction}
Hashkat is a simulation package designed to study the growth and time evolution of online social networks. As online social networks continue to grow in relevance, it has become increasingly important to quantitatively analyze their behaviour. We show that hashkat can be used to produce existing analytical graph models, as well as new, unstudied graphs with similar topologies to online social networks. Hashkat falls under  the field of agent based social simulations \cite{li2008agent}. Hashkat takes a kinetic Monte Carlo (kMC) approach, allowing a user to explore the characteristics of online social networks as they evolve through time. A wealth of recent work \cite{gleeson2014competition, brach2014spreading, myers2014bursty, doerr2011social} has focused on analyzing the topology of online social networks, in order to make inferences about information flow, interaction mechanisms, and network stability. Brach et al. \cite{brach2014spreading} were able to make predictions about the behaviour of information diffusion within social networks based on a random network topology. The model they developed described the evolution of rumours on social networks and gave qualitatively good results with respect to how messages propagate within Twitter.  Gleeson et al. \cite{gleeson2014competition} focused on meme diffusion for a directed social network. They found that the popularity growth of each meme can be described by a critical branching process. The popularity distributions of the memes had heavy tails similar to the distribution of links on the internet \cite{albert1999internet} as well many degree distributions for online social networks \cite{mislove2007measurement, ugander2011anatomy, java2007we}.

Currently, there are several commercial and non-commercial simulation packages available that simulate some aspect of online social networks. SMSim \cite{de2013simulation} simulates the network surrounding one agent and studies how information propagates from the central agent throughout the local neighbourhood.  SeSAm \cite{klugl1998multi} is able to treat a moderately small number of agents (tens of thousands) with an agent based modelling approach. Zeng \textit{et al.} \cite{Zeng2015} discussed a simulation method where random sampling techniques were used to build the initial network. Once the initial network was constructed, they then used a `close degree algorithm' to obtain snapshots of the social network at certain points in time. An R package called NetSim \cite{stadtfeld2013netsim} allows users to simulate the co-evolution of social networks and individual attributes. The engine of this package uses a generic Markov model, and the simulation incorporates social pressure (a user will connect to friends of a user they are connected to) and likeness of agents when making connections. 

Hashkat represents a significant step forward in the area of simulating online social behaviour due to its combination of scalability and features. Despite the complexity of the numerous features integrated into hashkat, it simplifies the user defined modelling process to the determination of rates for a given system. If the rates of events in a system are known (and correct), hashkat will time evolve the system along the true trajectory.

In this report, we first describe the design and high level structure of the algorithms used within hashkat. This includes a description of all events that can occur in the model. We then use hashkat to reproduce existing analytical graph models as test cases. This includes constructing random and preferential attachment graphs. Lastly, we walk through example use cases for which no analytical solutions are known.

\section{Algorithms, design, and features of hashkat}

\subsection{Engine}
We begin by first discussing the core engine of hashkat. Prior to using the software, rates at which events occur within the network must be set. These include the rate of content generation/propagation (i.e. tweeting and retweeting), the rate of agents connecting (user following), and the rate of agent creation (users joining the network). These rates are used to evolve the simulation through time using the kMC algorithm \cite{voter2007introduction}. Such rates can be obtained from publicly available social network APIs. kMC, popular in molecular simulations, is a generic and highly scalable \cite{schulze2008efficient} algorithm used for generating event sequences based on input rates. It should be noted that kMC is only valid if the rates are correct, the events associated with the rates are Poisson type events, and the different events are independent of one another. The kMC algorithm is as follows:

\begin{program}
\BEGIN \\ %
\COMMENT{array of rates for $n$ events} 
  rates := [r_1, r_2, ..., r_n] 
  \WHILE simulation\_time < maximum\_time:
  \COMMENT{Get random number $u_1\in(0,1]$}
  u_1 := rand() 
  \COMMENT{Create binary tree for efficient event selection}
  binary\_tree = create\_binary\_tree(rates)
  R := binary\_tree.sum()
  event := binary\_tree.event\_select(u_1)
  carry\_out\_event(event)
  \COMMENT{Get random number $u_2\in(0,1]$, update simulation time and rates if needed}
  u_2 := rand()
  simulation\_time := simulation\_time -ln(u_2) / R
  update\_rates(rates)
  \END
\END
\end{program}


At each simulation  time step, a list of \textit{all} possible events is generated. From these events, a random choice is made among all possible outcomes. The probability of choosing a particular event is weighted based on how frequently (i.e. based on the defined rate) that event is expected to occur. Given a set of X events possible, we define the cumulative rate function as
\begin{equation}
R = \sum\nolimits_{x \in X} r(x).
\end{equation}
An event $x \in X$ occurring with $r(x)$ frequency has selection probability of $r(x) / R$. The time-step used is inversely proportional to the sum of the event rates in the system. In the standard kMC algorithm, two random variables, $u_1$, $u_2\in(0,1]$ are used for event selection and to progress through time respectively. After event execution, the kMC method advances the simulation time by 
\begin{equation}
dt = -\ln(u_2) / R
\end{equation}
As the number of events in the system gets arbitrarily large, over a time frame $t$, the expected value of an event $x$ will be $r(x)t$. To further simplify constructing the total rate within the simulations, we have applied the notion of homogeneous agent behaviour and collective events. Homogenous agent behaviour implies that agents within the same category will have equivalent rates associated with their behaviour. That is, two users with an identical categorization will rebroadcast a message with the same probability, and connect with other users with the same rate. Although the agents within the same category act identically, they may follow different agents diversifying information diffusion. Hashkat can be used with an arbitrary number of agent profiles, but for optimal performance, 10-100 agent profiles is recommended.

\subsection{Program structure}
\begin{figure*}[h!]
  \centering
  \includegraphics[width=\linewidth]{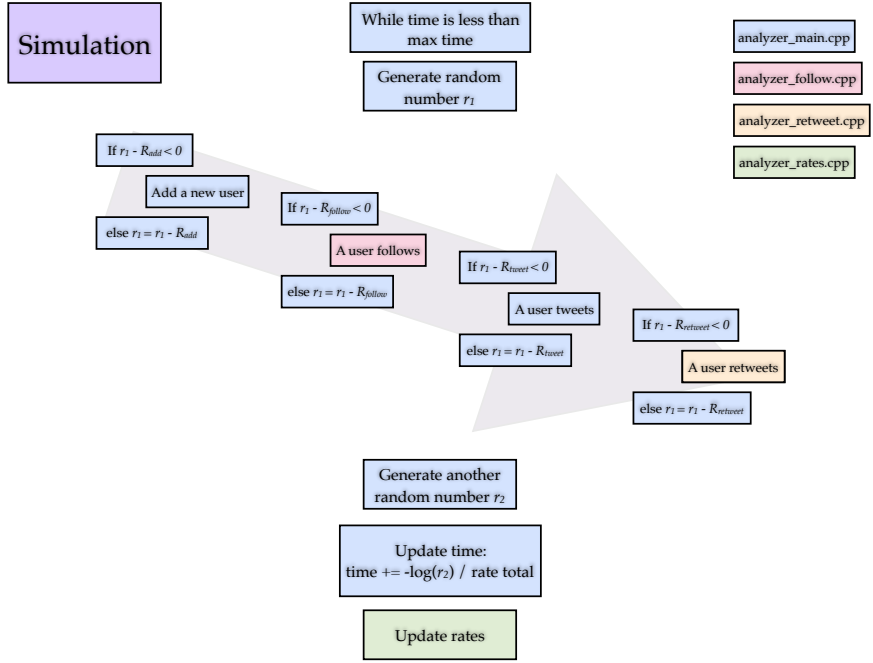}
  \caption{An illustration of kMC event selection with ``collective events'', as applied in hashkat.}
  \label{fig:code_map}
\end{figure*}

Due to the use of the homogenous agents, a collection of identical events can be treated as a single event when constructing the cumulative rate function (collective events). This can be seen in Figure \ref{fig:code_map}. Every collective event summarizes the total rate of children events (possibly collective themselves), forming a tree. Furthermore, collective events that only require selecting an agent from a set of homogeneous agents can be fulfilled in $O(1)$ (on average) time throughout our simulation. Our implementation first chooses between either the event categories of information diffusion (message broadcasts and rebroadcasts), social graph growth (users joining), or an agent connection forming (user following). For new content broadcasting, and new followings, hashkat makes use of homogeneous agent behaviour to reason about large lists of similar users. For message rebroadcasting, hashkat utilizes binary trees to choose a piece of content during information diffusion.

\begin{figure*}[h!]
\centering
\includegraphics[width=\linewidth]{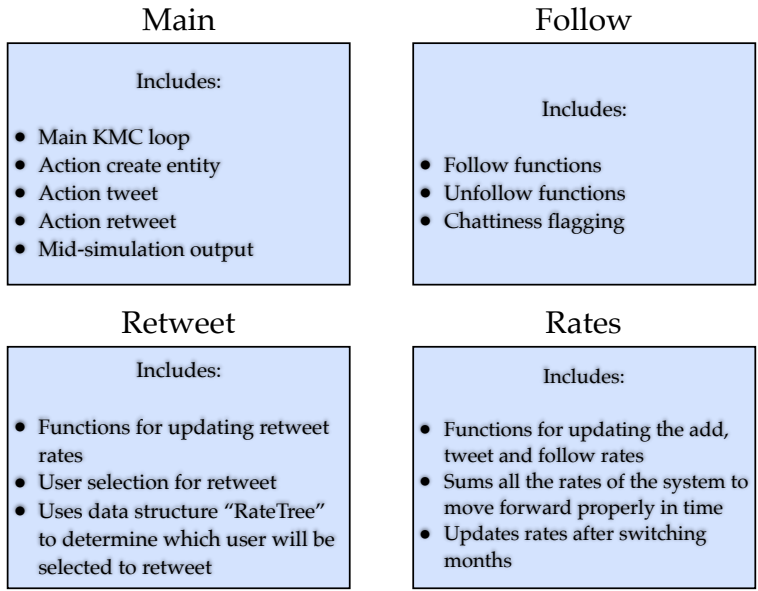}
\caption{A schematic view of the four most important classes used while a simulation is active.}
\label{fig:main_src_files}
\end{figure*}

Within the code base, there are four important C++ classes that should be mentioned. These can be seen in Figure \ref{fig:main_src_files}. These classes were used to separate important parts of the simulation into different files, therefore organizing the code base. As briefly explained in Figure \ref{fig:main_src_files}, the Analyzer class (analyzer\_main.cpp) is where the core of the simulation takes place. Here, the simulation either begins or continues (if a restart file was written to disk, i.e. serialized), all of the event functions are called, and statistics are output to monitor the simulation. Since some events are more complex than others (i.e. following and retweeting is more complex than tweeting) we have created separate classes for following and retweeting. These classes are AnalyzerFollow (analyzer\_follow.cpp) and AnalyzerRetweet (analyzer\_retweet.cpp) respectively. All of the connection functions are held within AnalyzerFollow. This includes functions to handle creating and destroying connections between agents. Also included here is the flagging of agents who may appear to send messages too often. This information could be used at a later time in the simulation for unfollowing (destroying a connection between agents). The AnalyzerRetweet class contains all functions and data structures for handling information propagation in the network. These functions include updating the cumulative retweet rate (which contributes to the cumulative rate function), and agent selection for a retweet event. The data structure used to organize agents is also here; this structure allows for an efficient agent selection when a retweet event occurs. Lastly, the AnalyzerRates (analyzer\_rates.cpp) class handles building the cumulative rate used in the kMC algorithm within the Analyzer class. 

\subsection{Simulation work flow}
\begin{figure*}[h!]
\centering
\includegraphics[width=\linewidth]{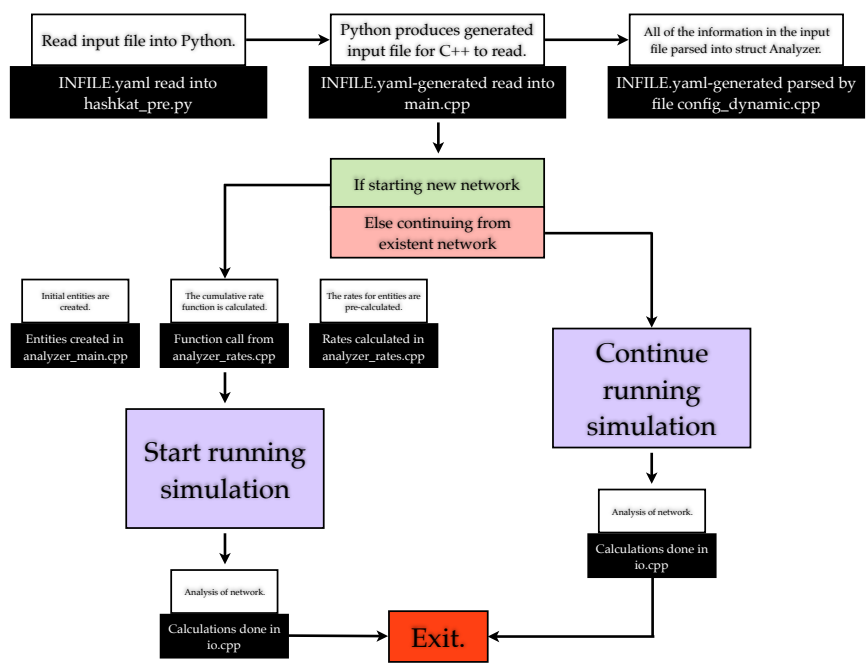}
\caption{The simulation work flow of hashkat. The white boxes are a high level description of the process and black boxes indicate which parts of the code are called.}
\label{fig:simulation_workflow}
\end{figure*}

To use hashkat (Figure \ref{fig:simulation_workflow}), a user starts by modifying an input file (INFILE.yaml). Here, the user can modify, add, or remove variables for their simulation. Once the user is satisfied with their simulation configuration, they execute a Bash script (run.sh) in the top level directory of hashkat. This first calls a Python script (hashkat\_pre.py) which generates other files needed for a simulation. As an example, the Python script generates the numerical grid from a user defined probability density function used in the retweeting algorithm. Once the python script has finished, the C++ executable is called. This executable first reads in the files generated by hashkat\_pre.py where the configuration information is stored within a class called AnalysisState (analyzer.cpp). If the user has a restart network file (from a previous run), and has selected the restart option then the existing network will be loaded into memory and will continue running. If not, initial agents are created, memory is declared, the cumulative rate function is calculated, time dependent rates are calculated for the simulation time duration, and the main kMC loop is entered. Once the simulation has concluded (either a maximum wall clock time or simulation time has been reached), basic analysis is done on the network (io.cpp) and the program exits. The analysis includes summary statistics of the network, degree distributions, and visualization files (all of which are located in the output directory).

\subsection{Agents}
\begin{figure}[h!]
\centering
\includegraphics[width=\linewidth]{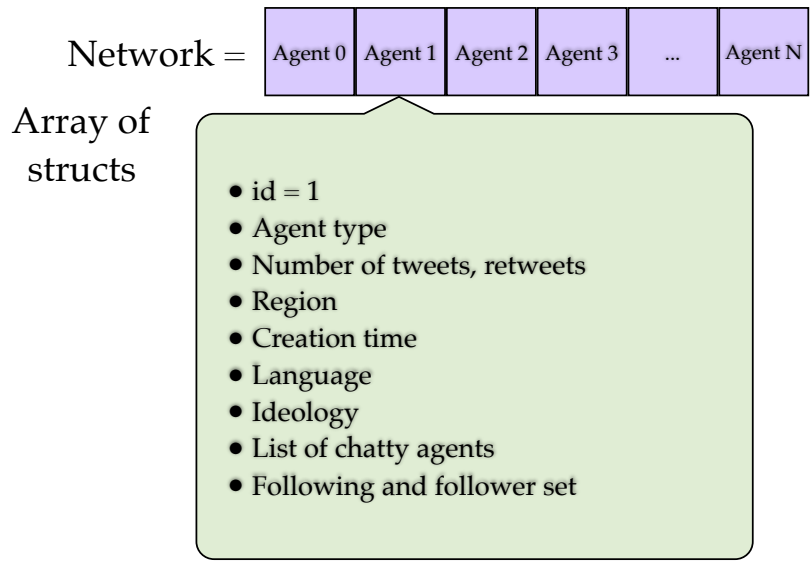}
\caption{A schematic view of the network built in hashkat outlining the attributes of agents within hashkat. Each agent has attributes which define both their content preferences and their generation profiles.}
\label{fig:agent_info}
\end{figure}

\begin{figure}[h!]
\centering
\includegraphics[width=\linewidth]{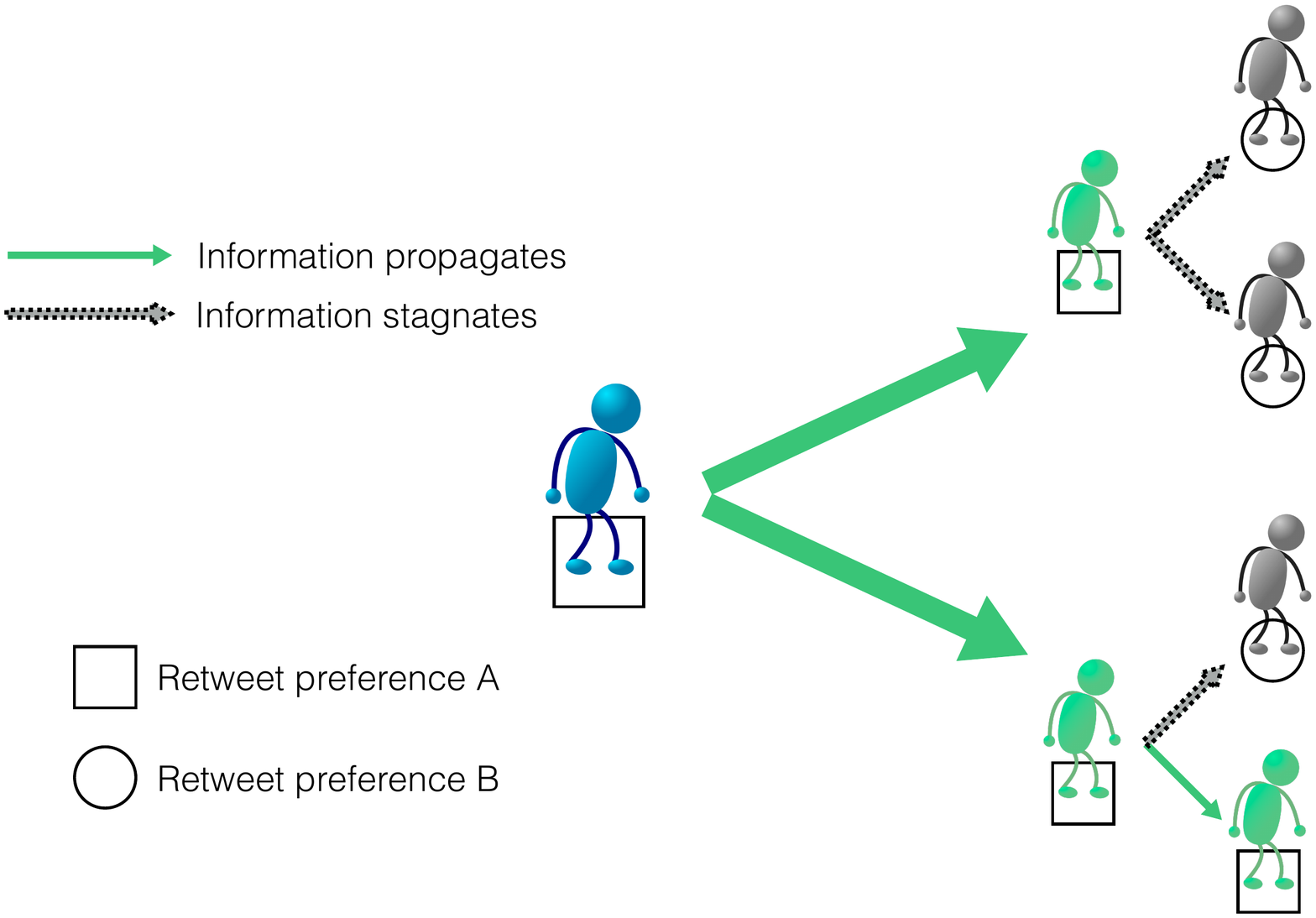}
\caption{A toy example outlining transmission factor in a network. In this example, users can either have a retweet preference for A (squares) or B (circles). This could be ideology, religion, or some other interest. The agent shown on the left sends out a message, and only agents where they have the same preference rebroadcasts the original message. The transmission probability is 1 for preference A and 0 for preference B. Although binary in this example, hashkat allows continuous transmission probability values when rebroadcasting.}
\label{fig:retweet_transmission}
\end{figure}

All agents within hashkat are classified by several attributes which describe their region, political ideology, language, humour preferences, and musical tastes. These attributes define both the semantic nature of content they create, as well as their reactions to other agent's content. These attributes can be seen in Figure \ref{fig:agent_info}. An agent's preferences for content discovery (i.e. finding like minded agents within the same region) are also determined by these attributes. Each of the attributes are discrete and have user defined weights associated with them. This allows for simulations with agents from multiple countries, speaking different languages, and even different personal preferences. These options allow for the possibility of creating networks for a wide variety of sociologically relevant cases. For example, a multilingual country with a wide variety of political beliefs can be described as a collection of multilingual agents with the same region attribute and a spread of political views. Conversely, agents within the same region may use different (incompatible) languages and exhibit strongly opposite political ideology. Such a simulation could describe a country where strife and civil discourse are possible. The structure of the graph generated under these two examples will be quite different. It should be noted that although we have used terms like ideology, humour, and music, these attributes can be mapped to arbitrary concepts or types of information. The important distinction between these labels is their scope, discoverability, and transmission factor. An example of transmission factor is shown in Figure \ref{fig:retweet_transmission}. In hashkat, messages transcend artificial barriers if certain conditions are met when comparing two agents. This can be seen in Table \ref{tab:boundaries}.

\begin{table}
\centering
\begin{tabular}{c|c}
Type of tweet & Rebroadcast condition\\ \hline
Political & Two agents share the same ideology, region, and language.\\
Humorous & Two agents share the same language.\\
Musical & No condition.\\
\end{tabular}
\caption{Table outlining how different tweets transcend artificial boundaries on online social networks.}
\label{tab:boundaries}
\end{table}

As an example of what kind of agent types can be created, we have constructed and visualized a network with four agent types (see Subsection \ref{ss:viz}, Figure \ref{fig:visualizations}). We have labelled these agent types standard, celebrity, organization, and bot. Their labels are based on their external attributes (i.e. attributes outside the network that define their behaviour within the network) and are motivated by Twitter. On Twitter, real life celebrities are more highly connected than other users, have a much higher in-degree than out-degree, and make up very little of the network. Real life organizations tend to have more connections than standard users, have a similar out and in degree, and also make up very little of the network. Lastly, standard users and bots are usually less connected, have varied behaviour, but make up most of the network. Bots are a phenomena on Twitter where the account is driven by scripts rather than a person. An example is an account that follows only verified users on Twitter (either celebrities or well know organizations). This bot follows 212 thousand users indicating that celebrities and well known organizations only make up 0.06\% of the Twitter population (Twitter has a reported 313 active million users in the second quarter of 2016). The distinction between a celebrity and standard agent can be understood based on how other agents react to them. Standard agents are judged based only on their observed activity within the network whereas celebrity agents are intrinsically attractive or persuasive irrespective of their ``in-simulation behaviour.''

\subsection{Trending topics}
Trending topics on Twitter tend to be synonymous with `hashtags'. Hashtags are keywords that begin with the `\#' character. Twitter parses the user generated content stream to find hashtags, and uses them to classify messages into trending topic lists. These topic lists allow for global information diffusion throughout the network. In hashkat, we model these global topic lists as circular buffers of fixed size. When a new message is added to a full buffer, old messages are discarded. The size of the buffer can be determined experimentally based on the lifetime of content relevancy. It is assumed that the effect of messages older than those in this list can be neglected. The probability of an agent adding a hashtag into their message can be set prior to a simulation. If an agent adds a hashtag to their message, they are added to the circular buffer. This allows for agents within the network to search for specific hashtags (network-wide content discovery) and find other agents with similar interests.

\subsection{Adding agents}
Throughout a simulation, it is possible for agents to continuously join the network at a rate set by the user in the configuration file (similar to a real online social community). The agent add rate is one of many rates that can be set prior to a simulation. All rates in hashkat can be set to vary throughout the simulation (this requires additional user input). This kind of variation of the input rates is analogous to the changing of rules in cellular automata \cite{wolfram1986theory}. Simple changes in the rules can form very different complex systems as time evolves. An example of a realistic agent add rate may be low in the early days of a network, increase as it gains in size, and eventually decrease as the set of possible users is exhausted. When hashkat performs an `add agent' event, it selects an agent type and region based on the configured input and creates a new agent from the particular region. The new agent's region may be used to decide other attributes depending on the user's direction. For example, an agent from a particular region may be more likely to speak one language than another, and may be more likely to hold a particular political view.

\subsection{Connecting agents} 
There are many ways for agents to connect within hashkat to mimic the complexity of relationships in real online social networks. In hashkat, the follow rates use the homogeneous agent assumption for users which have the same classification. These correspond to the expected rate that a given agent makes new connections in the network. When hashkat performs a `follow agent' event, it selects a follow model based on the user configuration. Several connection mechanisms have been integrated into hashkat; random, twitter suggest (preferential attachment), agent (selecting a certain agent type), agent type with nested preferential attachment, compatible content (likeness of agents), reciprocal connection, and a combination of all follow models.
\begin{figure}[h!]
  \centering
  \includegraphics[width=250pt]{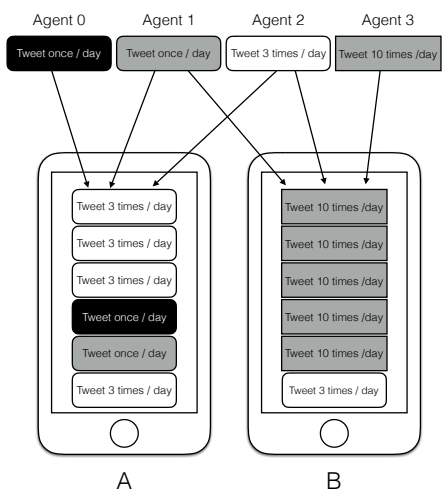}
\caption{An example of excessive tweeting and how it must be measured relative to a crowd. On screen A, agent 2 dominates the feed in comparison to agents 0 and 1. However, on screen B agent 3 dominates the feed. Even though agent 2 dominates the feed on screen A, they do not dominate the feed on screen B. Chattiness is in the eye of the beholder.}
\label{fig:chatty_example}
\end{figure}
Hashkat also allows for connections to be removed within the network (unfollowing). Currently, there are two unfollow mechanisms available. The first unfollow mechanism is random, and the second unfollow mechanism takes into account how often an agent tweets with respect to other agents that are being followed (excessive tweeting/chattiness). In hashkat, when an agent is followed, the tweet rate of the agent is compared with the average tweet rate of the agents previously followed. If the tweet rate exceeds the average, the newly followed agent is flagged and can be later unfollowed. Online social networks have the interesting property that subscribers may disagree about which agent is dominating their screen depending on the number and characteristics of agents they follow (Figure \ref{fig:chatty_example}). In this regard, discussions in an online setting can be quite different than what transpires in real life. For example, the overly talkative person at a dinner party is easy for everyone to identify; their rate of content generation is large compared to the other attendees. In an online setting, however, different users have different subscription lists; they are all effectively attending many dinner parties simultaneously. In practice, this means that while one observer may be overwhelmed by the content generated from a talkative user, to another observer, there may be a constant flow of messages from the rest of the subscription list making the talkative user seem less intrusive.\\

\subsection{Information propagation}
\begin{figure}[h!]
  \centering
  \includegraphics[width=250pt]{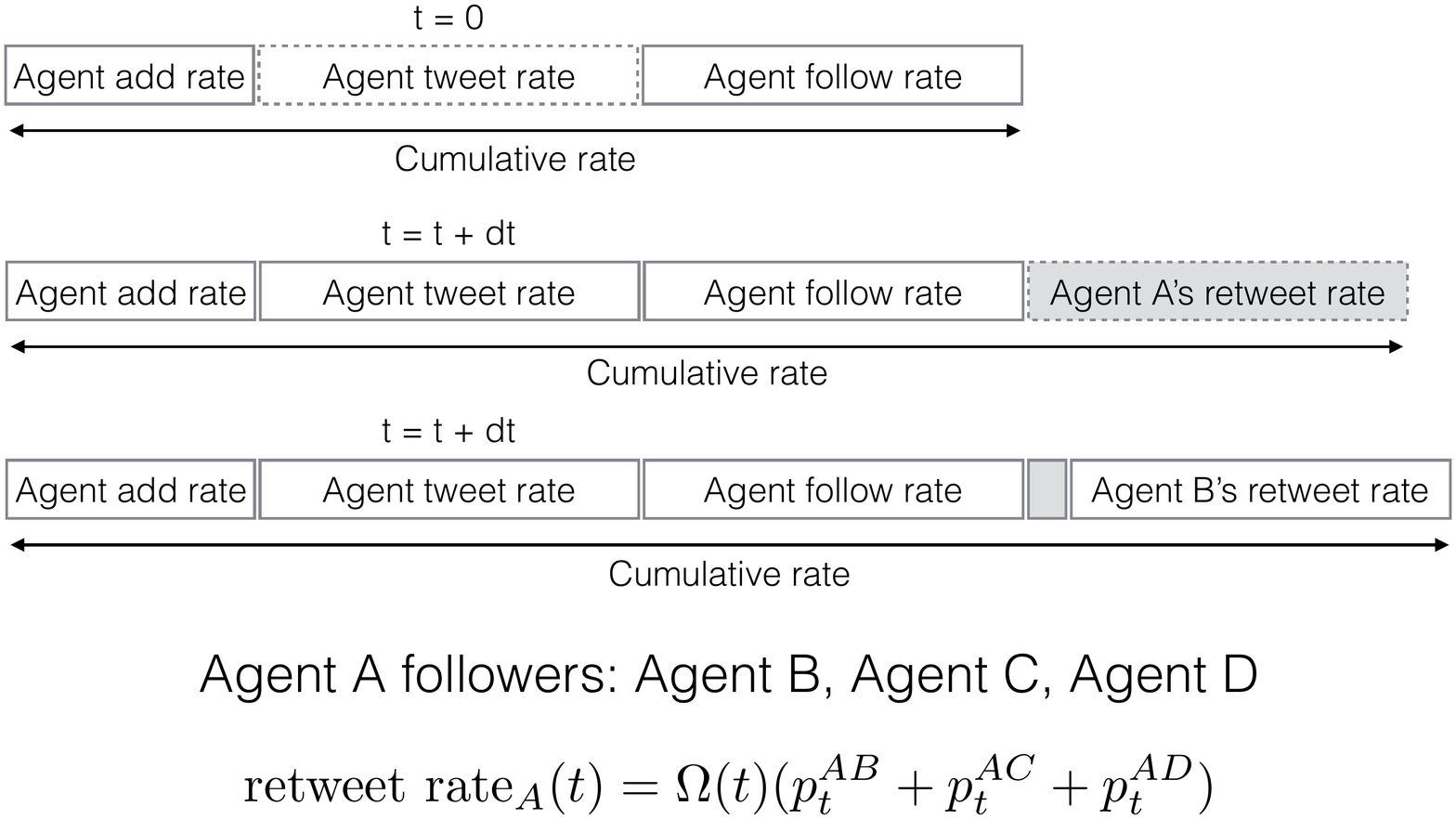}
  \caption{An example of how retweeting is incorporated into the cumulative rate function. At $t=0$, there are no current retweets and the agent tweet event is selected. Agent A is then selected and sends out a message. Moving forward in time, the retweet rate of agent A is first added into the cumulative rate function and then selected as the next event to be carried out. Moving forward in time once again, the retweet rate of agent A has decreased (due to $\Omega(t)$), agent B was selected to retweet, and therefore agent B's retweet rate is now added onto the cumulative rate function. Note that in the expression for the retweet rate of agent A, if $p_t = p_t^{AB} = p_t^{AC} = p_t^{AD}$, then the retweet rate becomes $\Omega(t)N_{\text{subscribers}}^Ap_t$ as seen in equation 3. Different functional forms of $\Omega(t)$ can be provided by the user. The consequences of this selection are explored in Subsection \ref{ss:viral}.}
  \label{fig:retweeting_example}
\end{figure}

One of the differentiating features of hashkat is the fact that agents within the simulation have the ability to create content (e.g. tweets or wall posts) in a topologically dynamic network. Moreover, other agents within the simulation have the ability to rebroadcast this content to their subscriber lists. The propensity for an agent to rebroadcast observed content depends both on the nature of the content and the receiving agent's own preference for such content. Different agents within the simulation can generate content with different meanings. If another agent agrees with the sentiment, they have a higher rebroadcasting probability. This probability can be any value on the interval [0,1]. To perform content rebroadcasts, we first make a selection between all active broadcasts. We select content $m$ with probability proportional to the collective rebroadcast rate $r(m)$ of the broadcasts audience. This rate depends on the makeup of the subscriber list. If subscribers agree with the content, the retransmission rate will be high. Otherwise, there will be a low probability of rebroadcasting. We consider the audience to be the subscribers of the broadcasting user $u_b$. The probability density function of rebroadcasting a given piece of content at a certain time after the content is created is an input to the system ($\Omega(t)$). The expected number of rebroadcasts can be calculated given $\Omega(t)$, and a transmission probability $\alpha$,
\begin{equation}
    N_{\text{rebroadcasts}}= \int \alpha N_{\text{subscribers}} \Omega(t)dt.
\label{eq:broadcast}
\end{equation}
Where $N_{\text{subscribers}}$ is the number of subscribers of the original broadcaster $u_b$. Note that the time-frame is reset for each rebroadcast and the rebroadcast is then treated in the same manner as an original broadcast. This allows for rebroadcasts themselves to be rebroadcast, analogous to how infections spread through a population. A schematic explanation of this can be seen in Figure \ref{fig:retweeting_example}. The Susceptable-Infected-Recovered (SIR) epidemic model \cite{pastor2015epidemic} is very similar to the mechanism we have applied to propagate messages. In this model, they consider the effective spreading rate of a disease to be $\lambda=\beta/ \mu$, where $\beta$ is the rate of infection and $\mu$ is the cure rate. In our model, $\mu=1$, and the effective spreading rate of a disease is therefore $\alpha$ from Equation \ref{eq:broadcast}. For a subscribing agent $u_i$, the transmission probability $\alpha$ is a piecewise function which depends both on $u_i$ and the broadcasting agent $u_b$. To facilitate this, the subscriber set of $u_b$ is partitioned into separate structures, each with their own summary data. These structures are implemented using Google's sparsehash \cite{silverstein2010google} data structure. Rebroadcasting is considered idempotent. After an agent has rebroadcast a given message once, further rebroadcasts by the same agent will have no effect. To this end, every piece of content is also associated with a set of agents who have previously broadcast the content. The set of active broadcasts is implemented as a binary tree containing rate summary information at every node, necessary to locate the message to select. Selection and insertion are $O(\log n)$, where $n$ is the number of live broadcasts. Periodic cleanup is performed on the set to remove broadcasts unlikely to be rebroadcast. Once a message has been chosen, an agent is chosen with a uniform random selection on the sparsehash data structure to rebroadcast the message. It should also be noted that in hashkat, every message broadcast has a unique identifier. This allows for hashkat to track a particular message through the network as it propagates. Hashkat logs the information related to this process to the output directory allowing the user to investigate information propagation in the network even further.
 
 \begin{figure}[h!]
  \centering
  \includegraphics[width=0.49\linewidth]{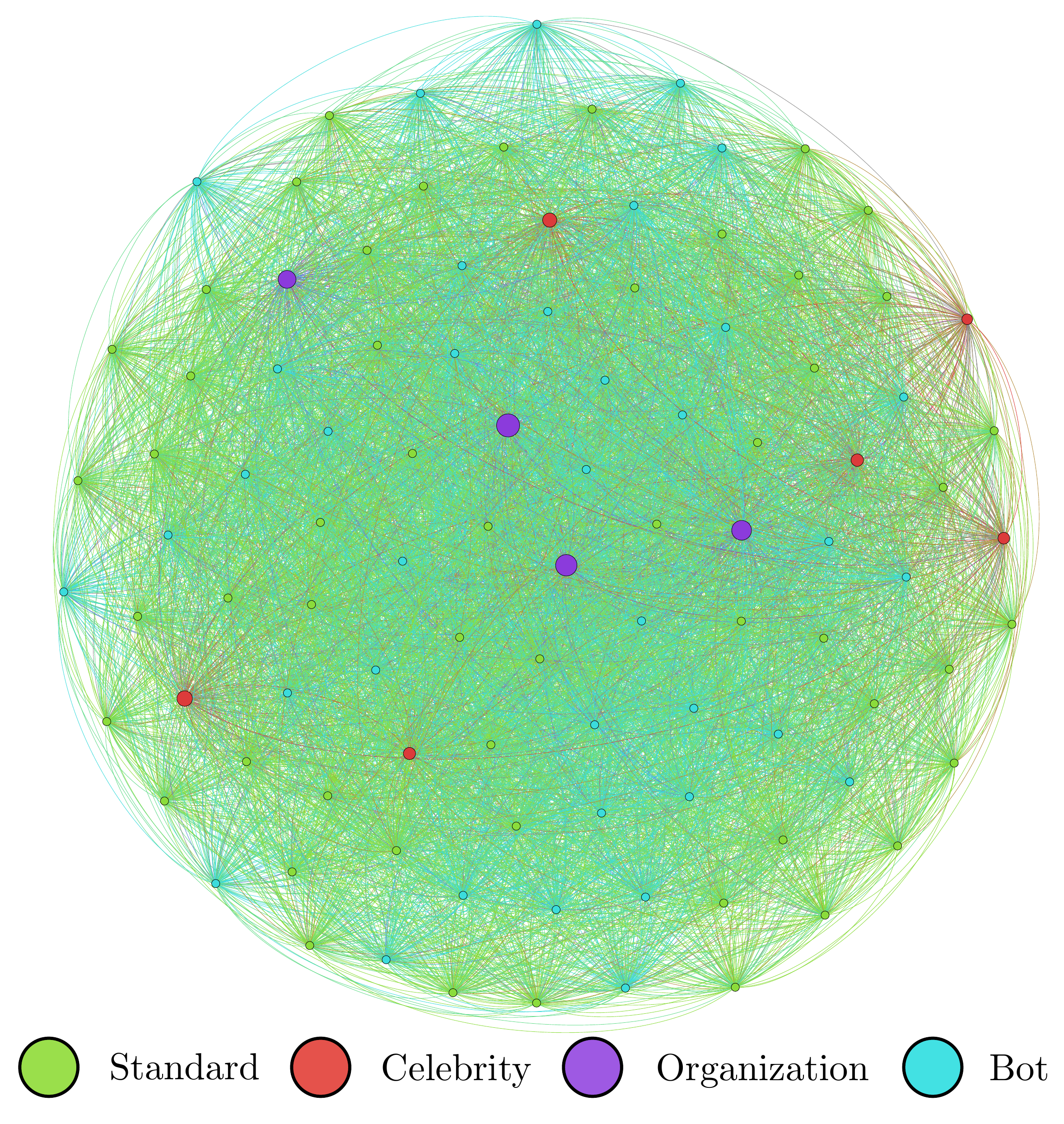}
  \includegraphics[width=0.49\linewidth]{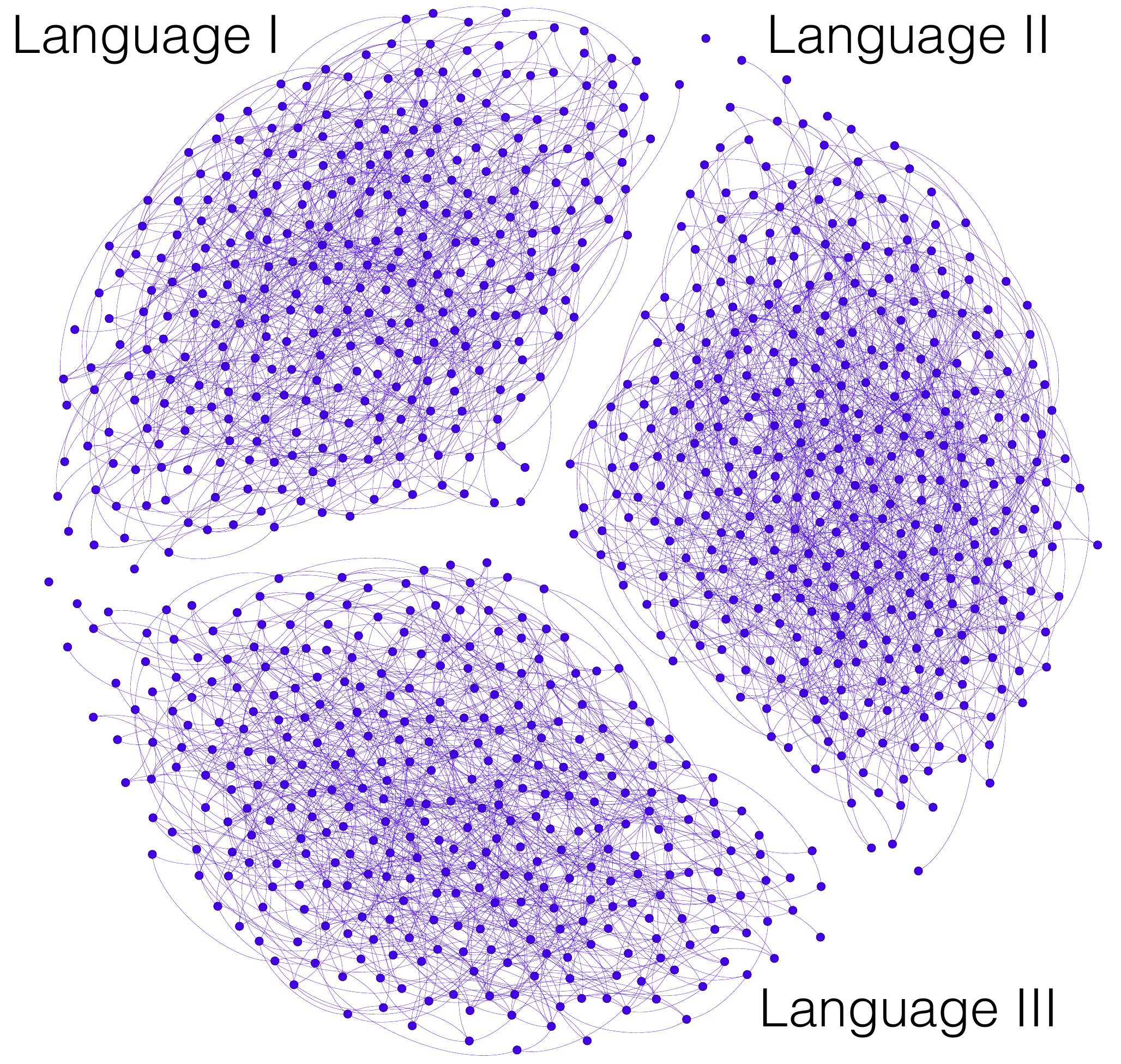}
  \caption{Hashkat is compatible with common network visualization and analysis tools (e.g. Gephi). On the left we show a random graph with different agent types (described in section 2.4). On the right, we show a random graph with one agent type where we have introduced 3 languages, causing 3 separate networks to emerge.}
  \label{fig:visualizations}
\end{figure}
 
\subsection{Visualization}
\label{ss:viz} 
\begin{figure}[h!]
  \centering
  \includegraphics[width=250pt]{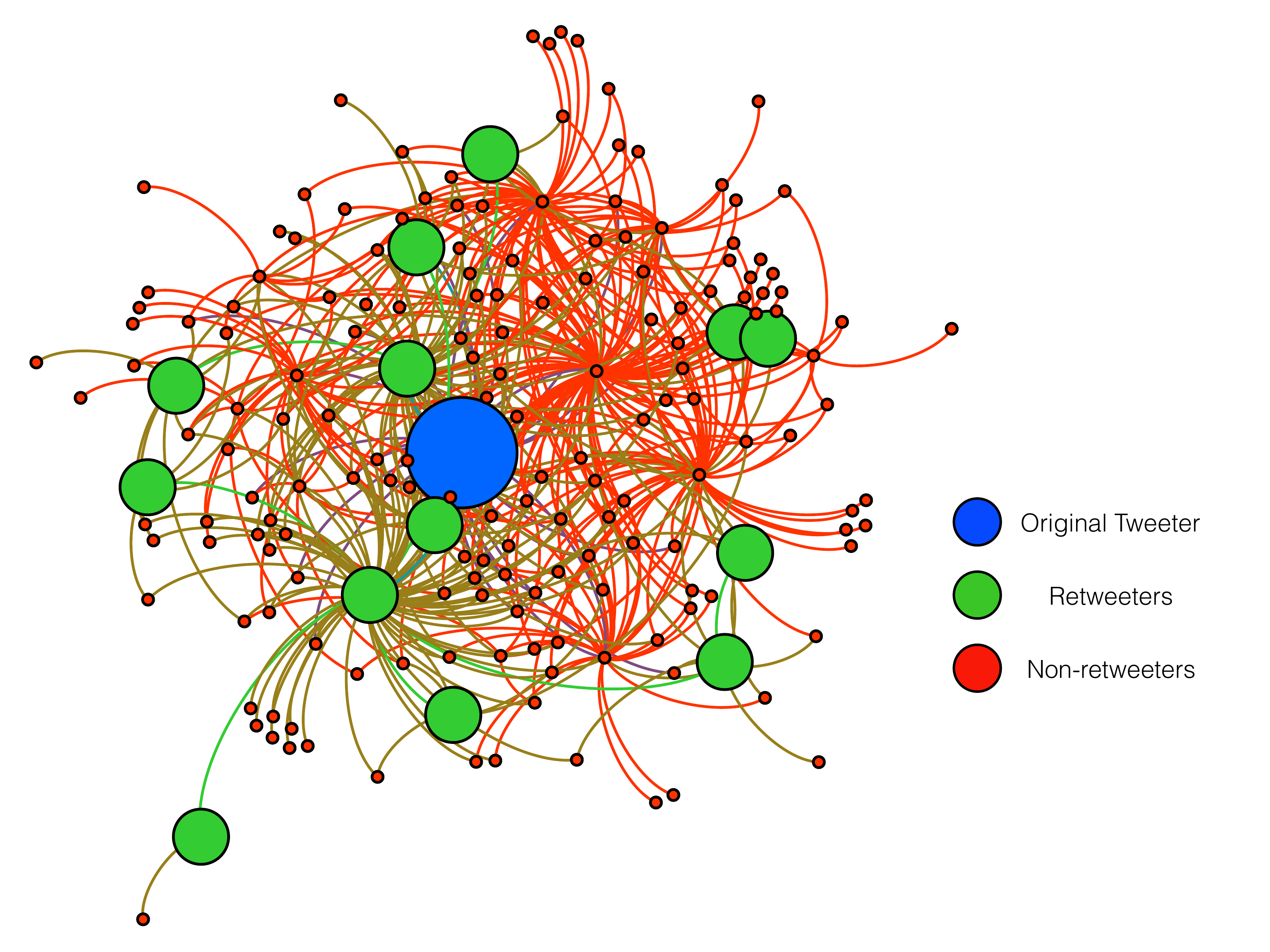}
  \caption{Visualization of the most rebroadcasted tweet. To create this visualization, hashkat tracks the tweet that has been most rebroadcasted within the simulation. Once the simulation concludes, all possible paths are drawn from the original broadcaster (largest node) down to all the rebroadcasters (medium sized nodes) and viewers (smallest nodes) of the original tweet.}
  \label{fig:retweet_viz}
\end{figure}

When a simulation has completed in hashkat, there are several different output files designed for visualizations. Some of these files are distributions including degree distributions, tweet count distributions, retweet count distributions, and network summary statistics. There are also two main files that can be used to visualize the network itself; one is designed for a network visualization tool called Gephi \cite{bastian2009gephi}, and the other can be used in Python's networkX \cite{hagberg2013networkx} package or R's iGraph package \cite{csardi2006igraph}. Two visualizations using Gephi of networks generated in hashkat can be seen in Figure \ref{fig:visualizations}. Apart from visualizing the full graph, hashkat is also capable of tracking the tweet that has been rebroadcasted the most, and creates a directed graph around that tweet. This can be seen in Figure \ref{fig:retweet_viz}.\\
  
\subsection{Compatibility and performance}
Hashkat is an open source project (GNU General Public License version 3) and has roughly 8600 lines of fully commented code. Hashkat has integrated build tests, unit tests, input tests, and has been built on all 3 major operating systems; Windows 10 (using the Linux subsystem), Linux, and OS X. It should also be noted that it has been run on systems that vary greatly in size. These includes systems with very small amounts of memory (Raspberry Pi) to systems with terabytes of memory (supercomputers). Scalability tests showed that hashkat can produce a random graph of 5 million nodes in roughly an hour, and 30 million nodes in approximately 40 hours. The project homepage (\href{http://hashkat.org}{http://hashkat.org}) has links to documentation (with an extensive set of examples), source code, and a web interface for sample input file generation.

\section{Test cases}
\subsection{Analytical graph models}
\begin{figure}[h!]
  \centering
  \includegraphics[width=250pt]{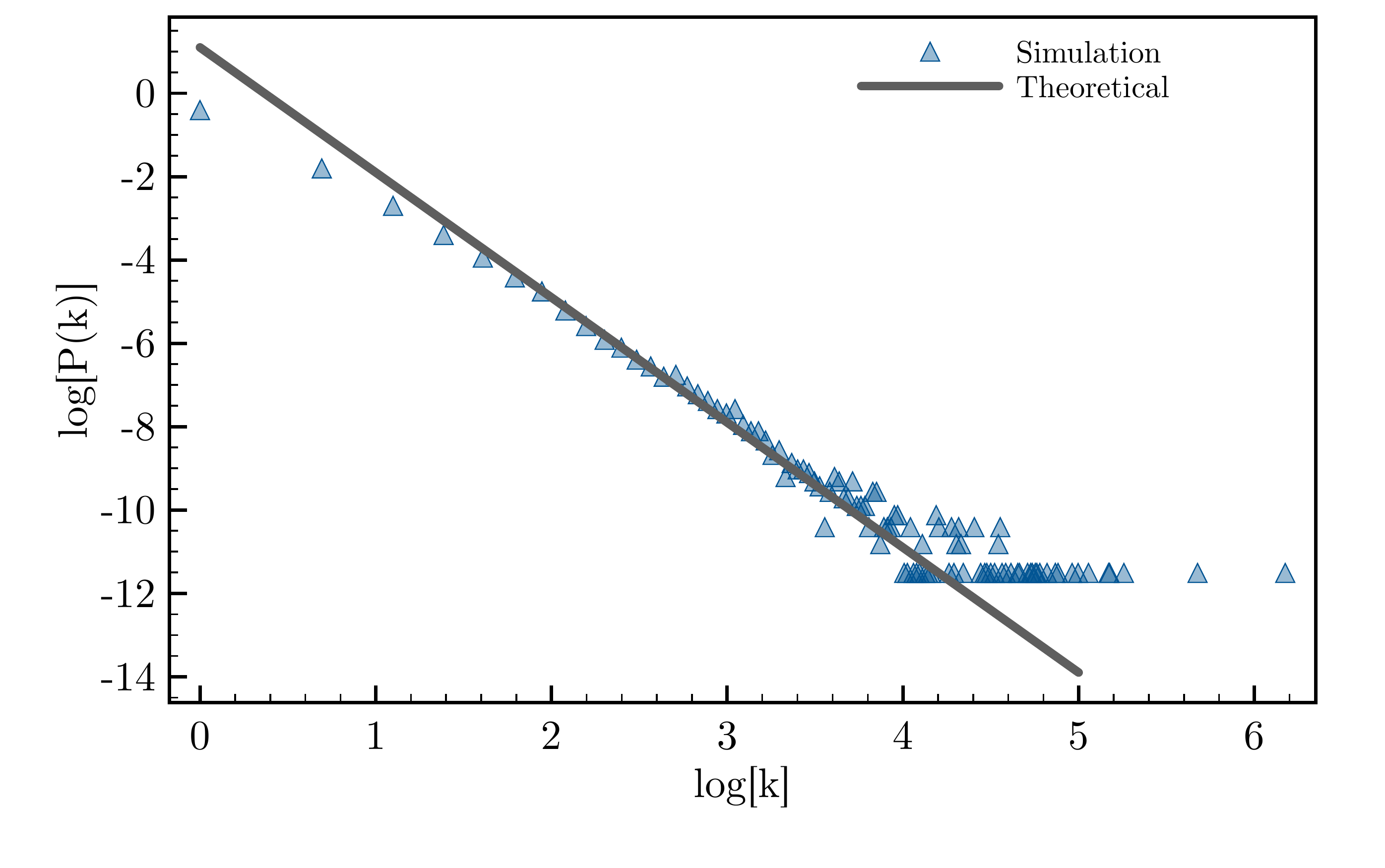}
  \caption{Comparison of the analytic degree distribution of the preferential attachment model to numerical results generated from hashkat. The points in the bottom right of the graph are a result of noise in the distribution.}
  \label{fig:pref_attach}
\end{figure}

\begin{figure}[h!]
  \centering
  \includegraphics[width=250pt]{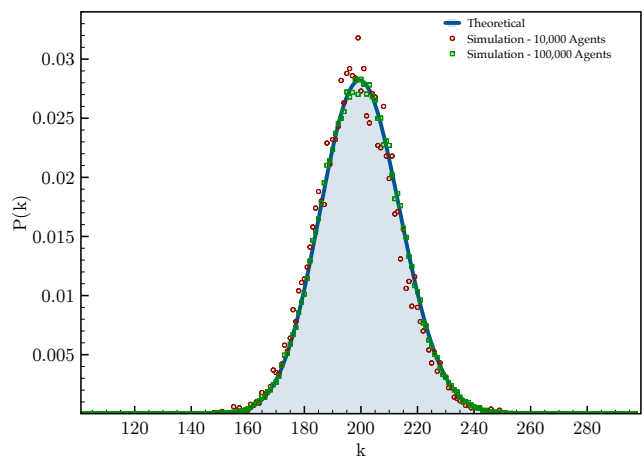}
  \caption{Comparison of the analytic degree distribution of a random graph to numerical results generated from hashkat.}
  \label{fig:random}
\end{figure}

We now use hashkat to firstly generate networks with topologies matching existing theoretical graph models. These models include the random graph, and the preferential attachment graph. These models are important because both have been solved analytically. As a result, their corresponding degree distributions can be expressed in closed form. This makes them an ideal test case for validating hashkat. For a random graph \cite{erdos1959random}, the degree distribution is known to be the binomial distribution
\begin{equation}
P(k)=\left(\begin{array}{c}
n-1\\
k
\end{array}\right)p^k(1-p)^{n-1-k},
\end{equation}
where $P(k)$ is the probability of finding a node with degree $k$. As $n$ (the total number of nodes) gets large, $P(k)$ can be approximated by the Poisson distribution
\begin{equation}
P(k)=\frac{\lambda^k e^{-\lambda}}{k!},
\end{equation}
where $\lambda = \langle k \rangle$.
When constructing a preferential attachment graph, the probability of creating a connection with a node $i$ is given by
\begin{equation}
P(k_i)=\frac{k_i}{\sum_{j=1}^{N}k_j},
\end{equation}
where $k_i$ is the degree of node $i$. In this model, the more connections a node has, the more probable the node is to obtain another connection. This is analogous to the concept of ``the rich get richer.'' Barab\'{a}si \textit{et al.} \cite{albert1999internet} have shown that the degree distributions for these graphs are
\begin{equation}
P(k)\approx k^{-\gamma},
\end{equation}
where $\gamma=3$ for large graphs. 
In Figures \ref{fig:pref_attach} and \ref{fig:random}, we compare our numerical results for both models with the the corresponding analytical degree distribution. This form of validation is a strong indication that hashkat is performing as designed. Note that the degree distributions with which we compare are not input to the simulation. 

\subsection{Reciprocal connections} 
\begin{figure}[h!]
  \centering
  \includegraphics[width=250pt]{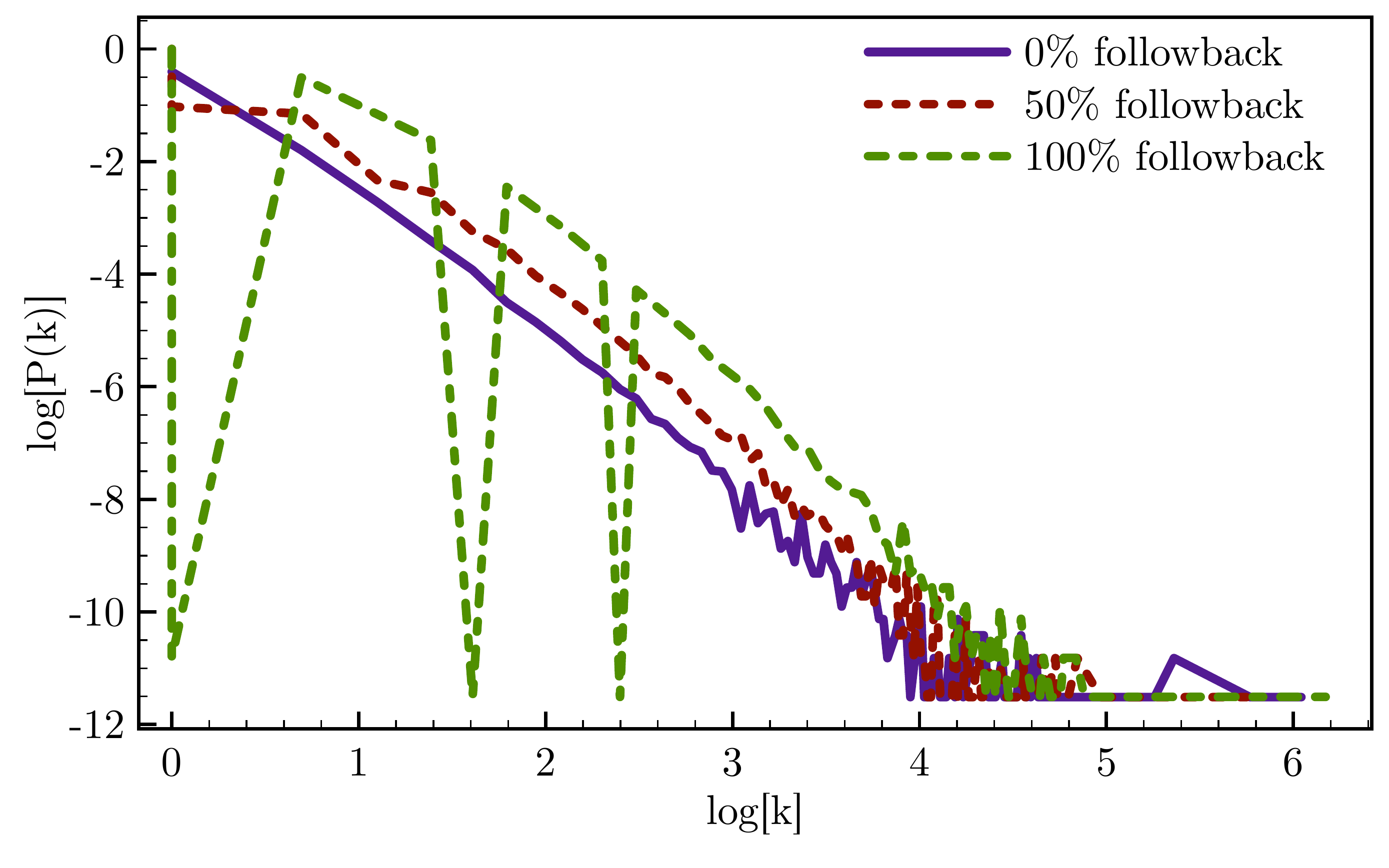}
  \caption{Degree distributions of three different simulations where the follow back probabilities were altered from 0-1 to show the effects of reciprocal links in a preferential attachment graph.}
  \label{fig:followback}
\end{figure}

Recently, a large-scale study of Twitter user profiles \cite{gleeson2014competition} indicated that despite the fact that the service is a directional graph, many of the relationships are in fact reciprocal. Frequently, when one user subscribes to another, there is a high probability of ``follow-back''. For Twitter, as many as 44\% of links are reciprocal. For demonstration purposes, we considered the impact that such a behaviour has on a preferential attachment network (i.e. where agents have the ability to see the global user list and preferentially follow high degree users). Hashkat can straightforwardly address this problem. We considered three cases: no follow-back, 50\% follow-back, and 100\% follow-back. All simulations were run using a constant agent add rate (to a total of 90,000 agents). The results shown in Figure \ref{fig:followback} are based on averages over 100 random initial seeds. As expected, the case of 0\% follow-back results in preferential attachment degree distribution. The 100\% results give a distribution which is non-zero only for even numbered degree. This is intuitive, as every subscription will necessarily generate a reciprocal link. The 50\% case (close to the real world Twitter observation), can be understood in terms of the these two extremes. Overall the distribution appears to be a preferential attachment one, but with the modification that the fraction of users with odd degree is less, whereas agents with even degree is enhanced. By considering a controlled system where the follow-back probability can be arbitrarily varied, it is possible to quantitatively observe the effect such user behaviour can have on network structure and time evolution. Note that this approach is substantially more straightforward than developing an extension to the analytic solution for the preferential attachment model, or attempting to gain this insight from experimental data directly, where only one realization of possible parameters can be observed.
 
\subsection{Viral content}
\label{ss:viral}
\begin{figure}[h!]
  \centering
  \includegraphics[width=\linewidth]{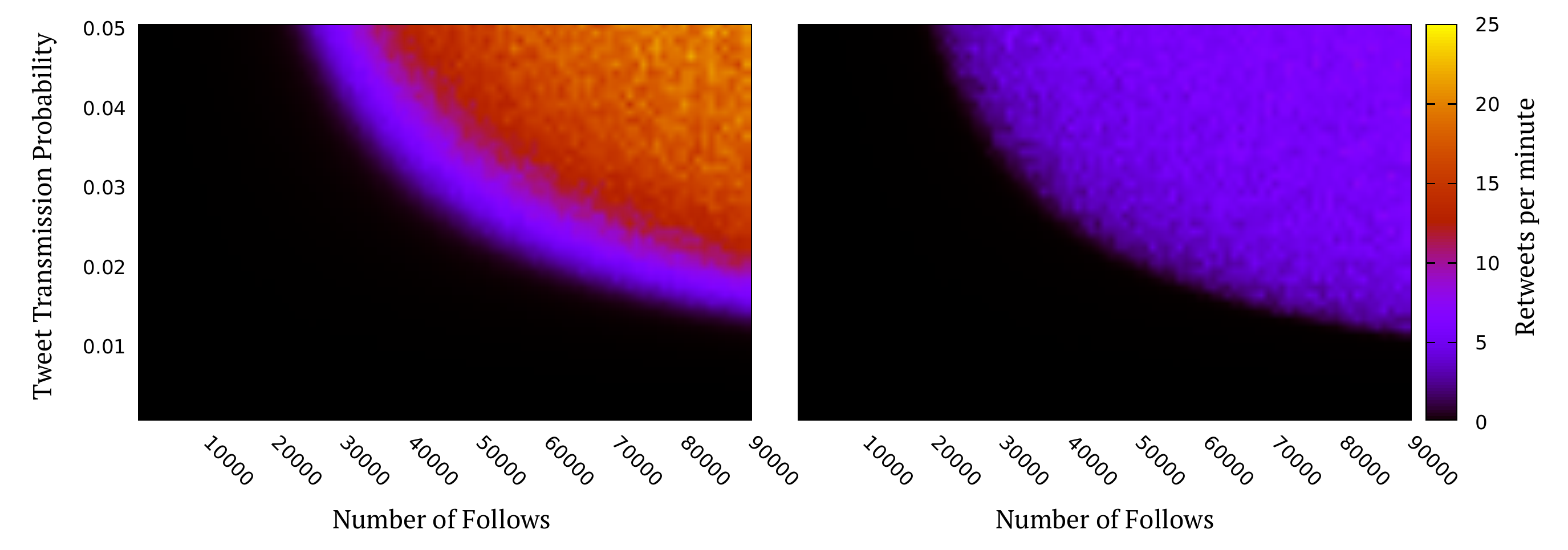}
  \caption{Contour plots showing distinct phases of content rebroadcast behaviour. The left plot has the probability density distribution (Equation \ref{eq:broadcast}) $\Omega(t)=t^{-1}$, and the right has $\Omega(t)=\exp(-t)$. }
  \label{fig:going_viral}
\end{figure}

Online social networks provide a medium for information to propagate. Information diffusion within these networks is of great interest in the literature \cite{gleeson2014competition, brach2014spreading, myers2014bursty, doerr2011social}. An interesting phenomena that emerges in large social networks is the concept of messages ``going viral''. This is a piece of information that is rebroadcast by many users over a short period of time. From equation \ref{eq:broadcast}, there are 3 variables that contribute to viral messages. Firstly, if an agent has many subscribers, the number of other agents who view their information is large. The content has a greater probability of being transmitted due to the audience size. Secondly, if the transmission probability is low, the message lifetime will be short. This is true even for agents with a large number of subscribers. The last determining factor is the probability density function giving the likelihood of a rebroadcast at some time after a message was broadcast. Initially, a message has a greater probability of being rebroadcast as it appears on the top of a user's message board (the feed). At some later time, the message gets pushed further and further down the feed due to newer messages decreasing the rebroadcast probability.

Using hashkat, we experimentally investigated the previously mentioned variables. To do so, we constructed dynamic random graphs with 1000 agents and varied the transmission probability ($0.001 \leq \alpha \leq 0.05$) as well as the probability density distribution $\Omega(t)$. For $\Omega(t)$, we used the functions $\Omega(t) = a\exp(-t)$ and $\Omega(t) = at^{-1}$ where $a$ was chosen such that the functions were normalized. These functions were then integrated over the interval $1\leq t\leq 600$ minutes. For each value of $\alpha$ ($\Delta\alpha = 0.001$) 10 simulations were run (with different seeds) and were averaged over. The agents in the simulations followed other agents at a constant rate until the total number of follows reached 90,000. 500 different simulations were run to produce the contour plots. By increasing the number of follows in the network, the average audience size increases. This allows for more rebroadcasts to occur. With a larger transmission probability, content becomes more likely to be passed on from one agent to another, increasing the effective lifetime of a message within the network. Looking at Figure \ref{fig:going_viral}, we notice a `phase transition' in the contour plots. When $\alpha \leq 0.01$ or the total number of follows $\leq 20,000$, no retweets occur. When $\alpha > 0.01$ or the total number of follows $> 20,000$, we see a transition in the contour plots. Travelling along the transition curve, the retweet rate in the network remains constant while the transmission probability and average degree varies. Any value of $\alpha$ or number of total follows above the transition curve allows for retweets to occur. When $\Omega(t)= at^{-1}$, more retweets occur as you move upwards and to the right of the transition curve. When $\Omega(t)=a\exp(-t)$, this is not the case. The exponential function approaches 0 quickly, meaning the probability of a retweet occuring at a time $t_{\text{retweet}} > t_{\text{tweet}}$ is much less than when $\Omega(t)=at^{-1}$. Despite this difference, the structure of the plots are similar for both probability density functions. Again, hashkat can straightforwardly and quantitatively explore information propagation while maintaining complexity.

\section{Conclusion}
Hashkat is a modern software tool designed for the study and simulation of online social networks. It is an agent based model with a diverse set of features and capabilities. Hashkat treats network growth and information flow simultaneously, allowing for users to study the interactions between these two phenomena. The kinetic Monte Carlo engine ensures an accurate time evolution of the system and requires only rates as an input. The tool is fully cross-platform (available on Linux, OS X, and Windows 10) and requires no commercial libraries or tools. Hashkat produces output which is compatible with existing social network analysis packages. The code can run on a wide variety of computing platforms. It can be used to accurately simulate networks ranging from simple random graphs to multi-million agent worlds with a variety of geographical regions, distinct languages, political views, and content preferences. The code is fully open source (GPL v3) and is freely available at the project homepage (http://hashkat.org). In addition to the source code and build instructions, the homepage has extensive documentation, sample visualizations, tutorials, and a web based tool to produce  sample input files. Hashkat is the most advanced simulation tool for online social networks in existence and is designed to enable fundamental research in this emerging platform of human interactions.

\begin{acknowledgements}
The authors would like to thank the National Sciences and Engineering Research Council of Canada (NSERC) for funding as well as Compute Canada for computational resources.
\end{acknowledgements}

\bibliographystyle{spphys}       
\bibliography{refs}{}   

\end{document}